\documentclass[twocolumn,twoside]{revtex4}
\usepackage{graphicx}
\usepackage{fancyhdr}
\def\dbar{\overline{D}{}^{\,0}}

\def\Dkskspp{D^0\ra K^0_S\,K^0_S\,\pi^+\pi^-}
\def\Dksksks{D^0\ra K^0_S\,K^0_S\,K^0_S}

\def\kskspp{K^0_S\,K^0_S\,\pi^+\pi^-}
\def\Dkspp{D^0\ra K^0_S\,\pi^+\pi^-}
\def\Dkk{D^0\ra K^+\,K^-}
\def\Dpp{D^0\ra\pi^+\pi^-}

\def\DstarDpi{D^{*+}\ra D^0\pi^+_s}
\def\KSpp{K^0_S\ra\pi^+\pi^-}

\def\cp{\textit{CP}}
\def\cpv{\textit{CPV}}
\def\Acp{A^{}_{CP}}
\def\acp{a_{CP}^{T}}
\def\ra{\!\rightarrow\!}
\def\Acpdet{A^{\rm{det}}_{CP}}
\def\Aslpi{A_{\epsilon}^{\pi_{s}}}
\def\Afb{A_{\rm FB}}
\def\Acpcor{A_{CP}^{\rm{cor}}}

\def\pt {p^{}_{\rm T}}
\def\costh {\cos\theta^{}_{\pi^{}_s}}
\def\cosths {\cos\theta^*}

\def\simge{\mathrel{%
   \rlap{\raise 0.511ex \hbox{$>$}}{\lower 0.511ex \hbox{$\sim$}}}}
\def\simle{\mathrel{
   \rlap{\raise 0.511ex \hbox{$<$}}{\lower 0.511ex \hbox{$\sim$}}}}

\usepackage{color}
\usepackage[dvipsnames]{xcolor}
\usepackage{comment}

\begin{document}

\title{\boldmath Measurement of the branching fraction and search for \cp\ violation in $\Dkskspp$ decays at Belle}

%

\author{Aman Sangal\\}
\affiliation{University of Cincinnati, Cincinnati, OH, USA, 45221}

\begin{abstract}
We measure the branching fraction for the Singly Cabibbo-suppressed decay 
$\Dkskspp$, and we search for \cp\ violation via a measurement of the 
\cp\ asymmetry $\Acp$ and also the $T$-odd 
triple-product asymmetry~$\acp$. The later two measurements are complementary. We use 922~fb$^{-1}$ of data recorded by the Belle experiment, which ran at the KEKB asymmetric-energy $e^+ e^-$ collider.
The branching fraction is measured relative to the Cabibbo-favored
normalization channel $\Dkspp$. Singly Cabibbo-suppressed charm decays are expected to have especially good sensitivity to new physics effects. 
\end{abstract}

\maketitle

\thispagestyle{fancy}


\section{Introduction}
Equal amounts of matter and antimatter
existed in the early Universe~\cite{Allahverdi_2021}.
For such an initial state to evolve into our current matter dominated universe \cite{Canetti:2012zc,Farrar:1993hn} violation of \cp\ (charge-conjugation and parity) symmetry~\cite{Sakharov:1967dj} is required.
The amount of \cp\ violation (\cpv) present in the Standard Model (SM) fails to account for the observed imbalance 
between matter and antimatter \cite{Farrar:1993hn,Huet:1994jb}. Thus, it is important to search for new sources 
of \cpv. 

In this note, we present the measurement of the branching fraction and search for \cpv\ in the singly Cabibbo-suppressed (SCS) decay $\Dkskspp$~\cite{charge-conjugates,Belle:2022gjv}. 
The branching fraction measurement gives an order of magnitude improvement in the precision over PDG world average \cite{Zyla:2020zbs}. In the SM framework, \cpv\ is expected to be very small ($\mathcal{O}(10^{-3})$ or smaller) in the charm meson decays \cite{Grossman:2006jg}. Any significant deviation from SM expectation will probe new physics effects beyond the SM. SCS decays are expected to be especially sensitive to physics beyond the SM, as their amplitudes receive contributions 
from QCD ``penguin'' operators and also chromomagnetic 
dipole operators~\cite{Grossman:2006jg}. 
The SCS decays $\Dkk$ and $\Dpp$~\cite{LHCb:2019hro}
are the only decay modes in which \cpv\ has been 
observed in the charm sector. 
The \cp\ asymmetry measured,
\begin{eqnarray}
\Acp & \equiv & \frac{\Gamma(D^0\ra f) - \Gamma(\dbar\ra \bar{f})}
            {\Gamma(D^0\ra f) + \Gamma(\dbar\ra \bar{f})}\,,
\end{eqnarray}
is small, at the level of~0.1\%.

For our analysis of $\Dkskspp$ decays, we search for \cpv\ in two 
complementary ways. We first measure the asymmetry $\Acp$; a nonzero 
value results from interference between contributing
decay amplitudes. The \cp-violating interference term 
is proportional to $\sin(\phi)\sin(\delta)$ for $D^0$ decays,
where $\phi$ and $\delta$ are the weak and strong
phase differences, respectively, between the amplitudes. Thus, to observe $\Acp\neq 0$, $\delta$ must be nonzero.

To avoid the need for $\delta\neq 0$, we also search for 
\cpv\ by measuring the asymmetry in the triple-product 
$C^{}_{T} = \vec{p}^{}_{K^0_S} \cdot (\vec{p}^{}_{\pi^+} \times \vec{p}^{}_{\pi^-})$,
where $\vec{p}^{}_{K^0_S}$, $\vec{p}^{}_{\pi^+}$, and $\vec{p}^{}_{\pi^-}$ are
the three-momenta of the $K^0_S$, $\pi^+$, and $\pi^-$ 
daughters, defined in the $D^0$ rest frame. From the two 
$K^0_S$ in final state we choose the $K^0_S$  with the higher momentum for this calculation.
The asymmetry is defined as
\begin{eqnarray}
A^{}_T & \equiv &  \frac{N(C^{}_T\!>\!0) - N(C^{}_T\!<\!0)}
{N(C^{}_T\!>\!0) + N(C^{}_T\!<\!0)}\,,
\end{eqnarray}
where $N(C^{}_T\!>\!0)$ and $N(C^{}_T\!<\!0)$ correspond to the yields of $\Dkskspp$ 
decays having $C^{}_T\!>\!0$ and $C^{}_T\!<\!0$, respectively. 
For $\dbar$ decays, we define the analogous $\cp$ conjugate quantity
\begin{eqnarray}
\bar{A}^{}_T & \equiv &  \frac{\overline{N}(-\overline{C}^{}_T>0) - \overline{N}(-\overline{C}^{}_T<0)}
{\overline{N}(-\overline{C}^{}_T>0) + \overline{N}(-\overline{C}^{}_T<0)}\,,
\label{eqn:atbar}
\end{eqnarray}

The difference is a $\cp$ violating observable
\begin{eqnarray}
\acp & \equiv & \frac{A^{}_T - \bar{A}^{}_T}{2} 
\end{eqnarray}
proportional to $\sin(\phi)\sin(\delta) $\cite{Durieux:2015zwa,PhysRevD.39.3339,Bensalem:2000hq}, and, unlike $\Acp$,
$\delta=0$ results in the largest \cp\ asymmetry. The observable $\acp$ is advantageous to measure experimentally,
as any production or detection related non $\cp$ asymmetry contribution cancels out.

\section{Reconstruction and data sample}
To measure the branching fraction and search for $\cp$ violation, we use 922~fb$^{-1}$ data collected by the Belle experiment\cite{Abashian:2000cg} running at KEKB asymmetric-energy $e^{+} e^{-}$ collider~\cite{kekb}. We use Monte Carlo (MC) simulated events to optimize event selection criteria, calculate reconstruction efficiencies, and study sources of background.

We reconstruct the decay chain $\DstarDpi$, $\Dkskspp$ where the charge of slow pion ($\pi_{s}$) is used to tag the flavor of $D$ meson. We start the reconstruction by selecting charged tracks that originate near $e^+e^-$ interaction point (IP) by requiring that the impact parameter $\delta z$ of a track along the $z$ direction (anti-parallel to the $e^+$ beam) satisfies $|\delta z| < 5.0$~cm, and that the impact parameter transverse to the $z$ axis satisfies $\delta r < 2.0$~cm. Tracks are identified as $\pi^{\pm}$ using likelihood based particle identification utilizing information from Belle subdetectors ACC (Aerogol cherenkov counter), TOF (Time of flight counter) and CDC (Central drift chamber). $K_S^0$ are reconstructed as $\KSpp$ using standard Belle neutral network based method \cite{Belle:2022gjv}. The invariant mass of the two pions is required to satisfy $|M(\pi^{+}\pi^{-}) - m^{}_{K_S^{0}}|<0.010$~GeV/$c^2$, where $m^{}_{K^0_S}$ is the $K^0_S$ mass~\cite{Zyla:2020zbs}. This range corresponds to three standard deviations in the mass resolution.

After identifying $\pi^{\pm}$ and $K_S^{0}$ candidates, 
we reconstruct $D^{0}$ candidates by requiring that the 
four-body invariant mass $M(K^0_S\,K^0_S\pi^+\pi^-)\equiv M$ 
satisfy $1.810~{\rm GeV}/c^2 < M < 1.920~{\rm GeV}/c^2$. 
We remove $\Dksksks$ decays, which have the same 
final-state particles, by requiring $|M(\pi^{+}\pi^{-}) - m^{}_{K^0_S}|>0.010$~GeV/$c^2$. This criterion removes 96\% of these decays. To improve the mass resolution, we apply mass-constrained vertex fits for the $K_S^{0}$ candidates. To ensure the $D^{0}$ daughters originate from a common decay vertex we perform a vertex fit using the $\pi^\pm$ tracks and the momenta of the $K_S^0$ candidates; the resulting fit quality ($\chi^{2}$) must satisfy a loose requirement to ensure that the tracks and $K^0_S$ candidates are consistent with originating from a common decay vertex.

To reconstruct $\DstarDpi$ decays, we combine $D^0$ candidates with $\pi^+_s$ candidates and require that the mass difference $M(K^0_S\,K^0_S\pi^+\pi^-\pi^+_s) - M\equiv \Delta M$ be less than 0.15~GeV/$c^2$. To reduce the combinatorial backgorund and remove the $D^{*+}$ candidates coming from $B$ decays, we also require that the momentum
of the $D^{*+}$ candidate in the CM frame be greater than 2.5~GeV/$c$; A vertex fit is performed for $D^{*+}$, constraining the $D^{0}$ and $\pi^+_s$ to originate from the IP. We subsequently require $\sum(\chi^{2}/\rm{ndf}) <100$, where the sum runs over the two mass-constrained $K^0_S$ vertex fits, the $D^0$ vertex fit, and the IP-constrained $D^{*+}$ vertex fit, and ``ndf'' is the number of degrees of freedom in each fit. After applying all selection criteria, for events with
multiple $\DstarDpi,\,\Dkskspp$ signal candidates we retain a single candidate by choosing that with the lowest value of $\sum(\chi^{2}/\rm{ndf})$.

We measure the branching fraction relative to Cabbibbo favored $\Dkspp$ decays 
observed in the same data set. The branching fraction for $\Dkskspp$ is calculated as
\begin{eqnarray}
\mathcal{B}(\Dkskspp) & = &  \nonumber \\
 & & \hskip-1.5in 
\left(\frac{N_{K^0_S\,K^0_S\pi^+\pi^-}}{N_{K^0_S\pi^+\pi^-}}\right)
\left(\frac{\varepsilon^{}_{K^0_S\pi^+\pi^-}}{\varepsilon^{}_{K^0_S\,K^0_S\pi^+\pi^-}}\right)
\times \frac{\mathcal{B}(\Dkspp)}{\mathcal{B}(\KSpp)}\,, \nonumber \\
\label{eqn:br}
\end{eqnarray}
where $N$ is the fitted yield for $\Dkskspp$ or $\Dkspp$ 
decays;  
$\varepsilon$ is the corresponding reconstruction efficiency,
given that $K^0_S\ra\pi^+\pi^-$; 
and $\mathcal{B}(\KSpp)$ and $\mathcal{B}(\Dkspp)$ are the world average branching fractions 
for $\KSpp$ and $\Dkspp$~\cite{Zyla:2020zbs}.
The selection criteria for $\Dkspp$ are the same as 
those used for $\Dkskspp$, except that only one $K^0_S$ 
is required.

After applying all the selection criteria, the resulting reconstruction efficiencies are $\varepsilon^{}_{K^0_S\,K^0_S\pi^+\pi^-} = (6.92\pm 0.02)\%$ and $\varepsilon^{}_{K^0_S\pi^+\pi^-} = (14.97\pm 0.03)\%$, We calculate reconstruction efficiencies using MC simulation, so we consider data-MC correction factors of
$0.930\pm 0.014$ for $\Dkskspp$ and 
$0.899\pm 0.007$ for $\Dkspp$.

\section{Signal Extraction and branching fraction measurement}
\begin{figure}[h!]
\includegraphics[width=0.51\textwidth]{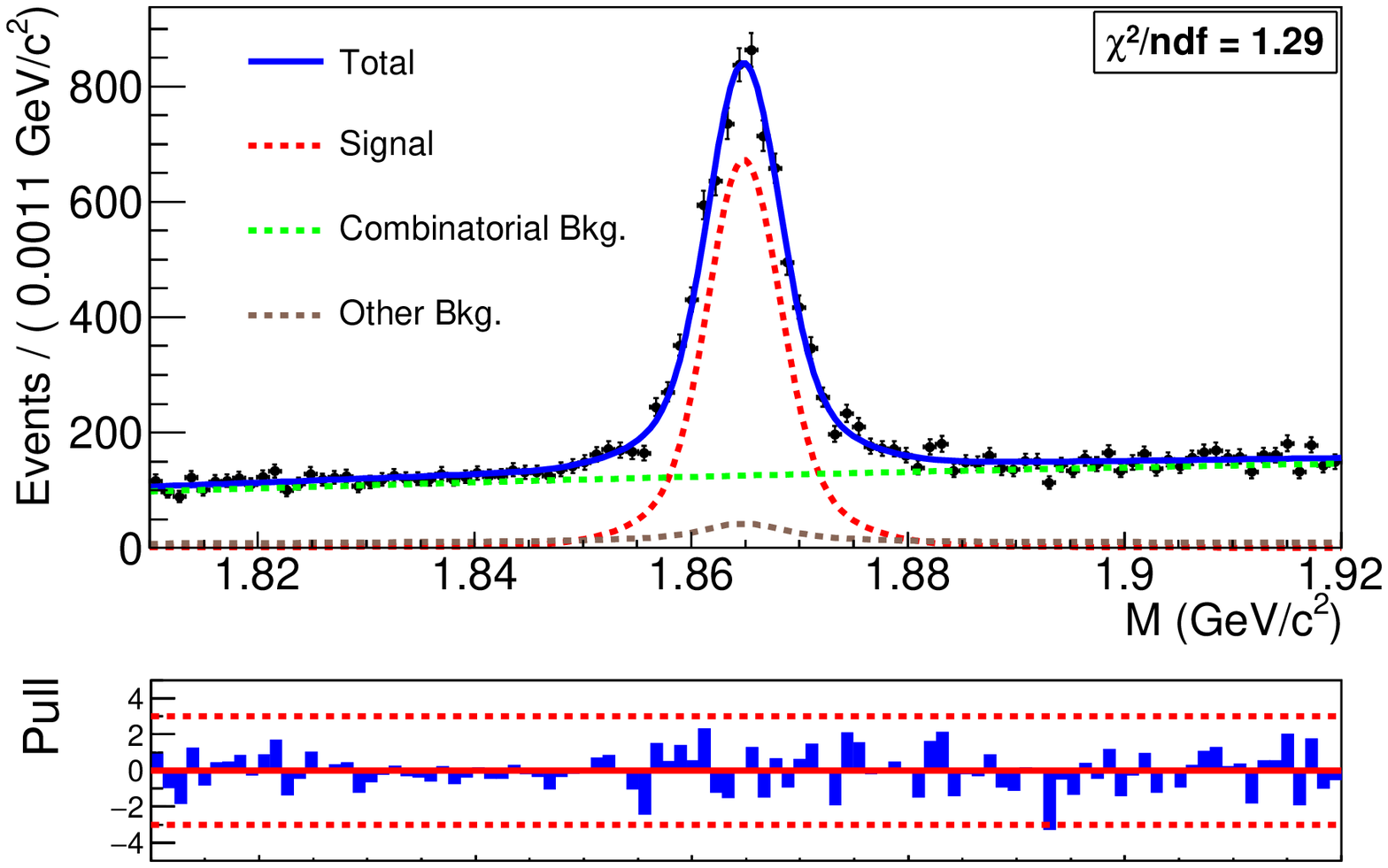}\\
\includegraphics[width=0.51\textwidth]{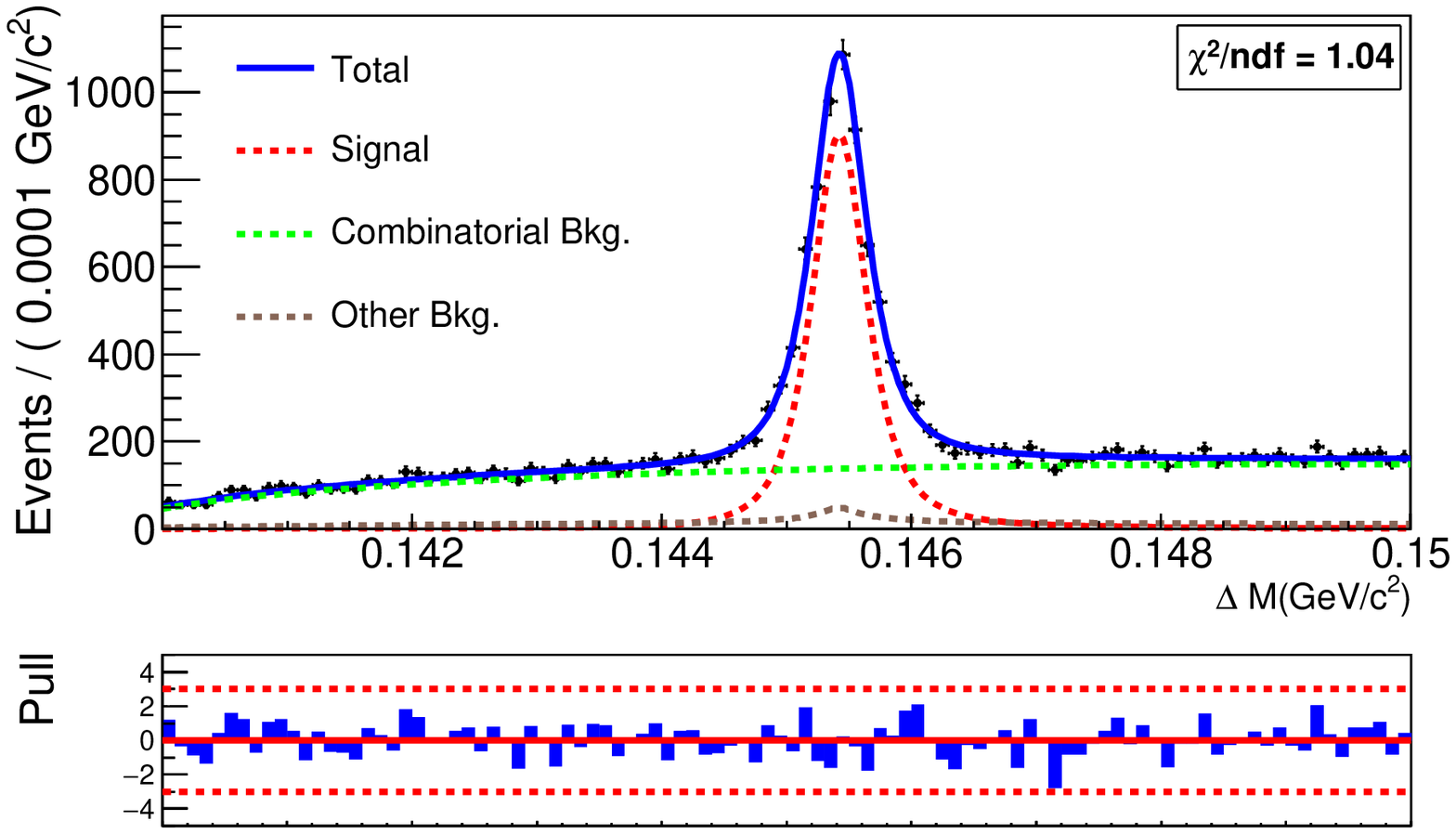}
\caption{Projections of the fit for $\Dkskspp$ on $M$ (upper) 
and $\Delta M$ (lower).
The brown dashed curve consists of
slow pion, broken charm, and $\Dksksks$ backgrounds.
The corresponding pull distributions 
[$= ({\rm data} - {\rm fit\ result})/({\rm data\ uncertainty})$] 
are shown below each projection. The dashed red lines correspond to $\pm 3\sigma$ values. 
}
\label{fig:BF_signal}
\end{figure}

For $\Dkskspp$, we determine the signal yield via a two-dimensional unbinned extended maximum-likelihood  fit to the variables $M$ and $\Delta M$. The fitted ranges are $1.810~{\rm GeV}/c^2 < M < 1.920~{\rm GeV}/c^2$ and $0.140~{\rm GeV}/c^2 < \Delta M < 0.150~{\rm GeV}/c^2$. The total sample was divided into the following categories of events: 
{\rm (a)}~correctly reconstructed signal events;
{\rm (b)}~mis-reconstructed signal events, i.e., one or more daughter tracks are missing;
{\rm (c)}~``slow pion background,'' i.e., a true $\Dkskspp$ decay is combined with an 
extraneous $\pi^+_s$ track;
{\rm (d)}~``broken charm background,'' i.e., a true $\DstarDpi$ decay is reconstructed, 
but the (non-signal) $D^0$ decay is mis-reconstructed, faking a $\Dkskspp$ decay; 
{\rm (e)}~purely combinatorial background, i.e., no true $D^{*+}$ or $D^0$ decay; and
{\rm (f)}~$\Dksksks$ decays that survive the $M(\pi^+\pi^-)$ veto. The fit yields $6095\pm 98$ signal events. Projections of the fit are shown in Fig.~\ref{fig:BF_signal}.

\begin{figure}[h!]
\centering
\includegraphics[width=0.51\textwidth]{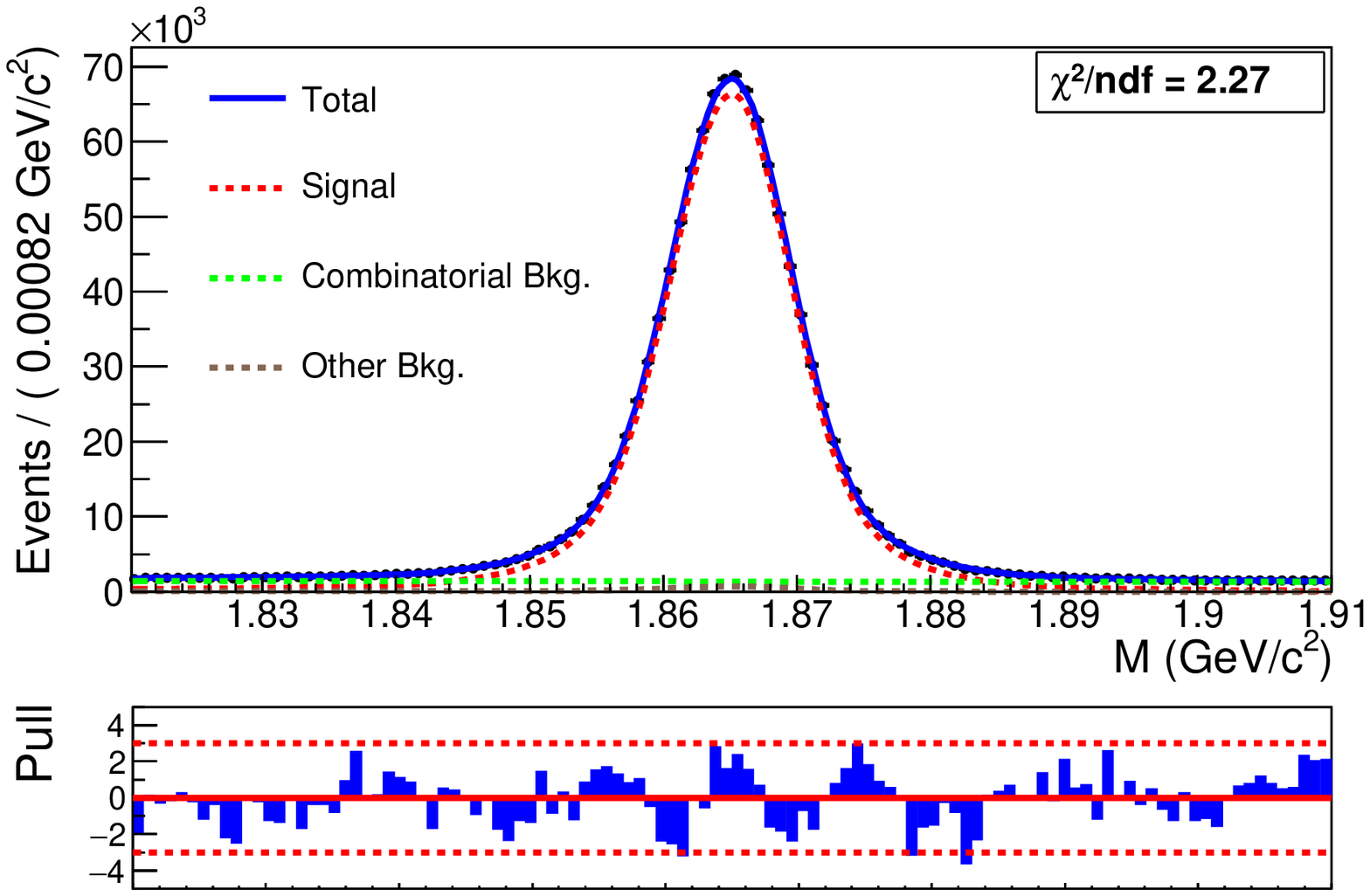}
\vspace{0.5cm}
\includegraphics[width=0.51\textwidth]{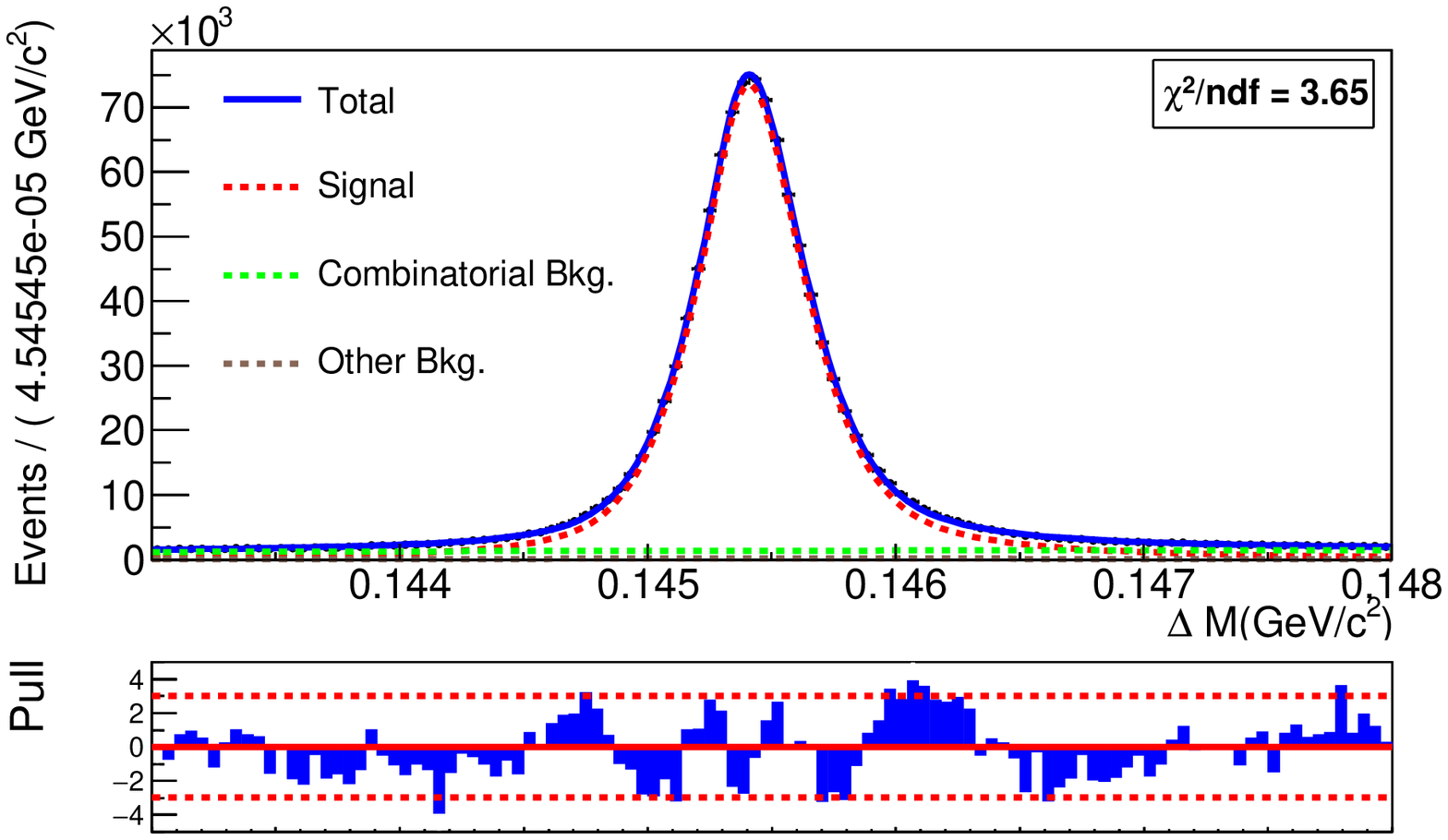}
\caption{Projections of the fit for $\Dkspp$ on $M$ (upper) and $\Delta M$ (lower).
The corresponding pull distributions 
[$= ({\rm data}-{\rm fit\ result})/{\rm data\ uncertainty})$] 
are shown below each projection. The dashed red lines correspond to $\pm 3\sigma$ values.
}
\label{fig:BF_norm}
\end{figure}

We determine $N^{}_{K^0_S\pi^+\pi^-}$ from a two-dimensional
binned fit (rather than unbinned, as the statistics are large) 
to the $M$ and $\Delta M$ distributions. The fitted ranges are 
$1.820~{\rm GeV}/c^2 < M < 1.910~{\rm GeV}/c^2$ and 
$0.143~{\rm GeV}/c^2 < \Delta M < 0.148~{\rm GeV}/c^2$~\cite{comment_range}. The fit yields $1\,069\,870\pm 1831$ $\Dkspp$ decays. 
Projections of the fit are shown in Fig.~\ref{fig:BF_norm}.\\
\indent Inserting all values  into Eq.~(\ref{eqn:br}), reconstruction efficiencies along with the fitted yields 
and the PDG values~\cite{Zyla:2020zbs} 
$\mathcal{B}(\Dkspp) = (2.80\pm 0.18)\%$ and
$\mathcal{B}(\KSpp) = (69.20\pm 0.05)\%$ gives 
$\mathcal{B}(\Dkskspp) = [4.82 \pm 0.08\,({\rm stat}) \, ^{+0.10}_{-0.11}\,({\rm syst}) \pm 0.31\,({\rm norm})]\times 10^{-4}$, where the first uncertainty is statistical, the second is systematic \cite{Belle:2022gjv}, and the third is from uncertainty in the normalization channel.

\section{Measurement of $\Acp$}
We measure the \cp\ asymmetry $\Acp$ from the difference 
in signal yields for $D^0$ and $\dbar$ decays:
\begin{equation} 
\Acpdet{} = \frac{N(D^{0} \ra f) - N(\dbar\ra \overline{f})}
                 {N(D^{0} \ra f) + N(\dbar\ra \overline{f})}\,.
\label{eqn:acp_def}
\end{equation}
The observable $\Acpdet$ includes asymmetries in production and reconstruction:
\begin{eqnarray}
\Acpdet & = & \Acp  + \Afb + \Aslpi\,,
\end{eqnarray}
where $\Afb$ is the ``forward-backward'' production asymmetry~\cite{Afb} between 
$D^{*+}$ and $D^{*-}$ due to $\gamma^{*}-Z^{0}$ interference in $e^{+}e^{-}\ra c\overline{c}$\,; 
and $\Aslpi$ is the asymmetry in reconstruction efficiencies for $\pi^\pm_{s}$ tracks. 

We correct for $\Aslpi$ in $\kskspp$ events by separately weighting $D^0$ and $\dbar$ decays:
\begin{eqnarray}
w^{}_{D^0}   & = & 1 - \Aslpi (\pt,\,\costh) \\
w^{}_{\dbar} & = & 1 + \Aslpi (\pt,\,\costh)\,.
\end{eqnarray}
where $\Aslpi$ is the asymmetry in $\pi_{s}$ detection in bins of $\pt$ and $\costh$ of 
the $\pi^\pm_s$, where $\pt$ is the transverse momentum and
$\theta^{}_{\pi^{}_s}$ is the polar angle with respect to the $z$-axis, both evaluated in the laboratory frame.

After correcting for $\Aslpi$, we obtain $\Acpcor = \Acp + \Afb$. The asymmetry $\Afb$ is an 
odd function of $\cosths$, and $\Acp$ is an even function, where $\theta^*$ is the polar angle 
between the $D^{*\pm}$ momentum and the $+z$ axis in the CM frame. We thus extract 
$\Acp$ and $\Afb$ via
\begin{eqnarray}
\Acp  & =  & \frac{\Acpcor(\cosths) + \Acpcor(-\cosths) }{2} \label{eqn:acp} \\
\Afb  & =  & \frac{\Acpcor(\cosths) - \Acpcor(-\cosths) }{2}\,. \label{eqn:nfb}
\end{eqnarray}
We calculate $\Acpcor$ in four bins of $\cosths$: 
$(-1.0,-0.4)$, $(-0.4, 0)$, $(0, 0.4)$ and $(0.4,1.0)$.
We determine $\Acpcor$ for each bin by simultaneously fitting
for $D^0$ and $\dbar$ signal yields for weighted events in that bin. The results for $\Acpcor$ are combined according to Eqs.~(\ref{eqn:acp}) and (\ref{eqn:nfb}) to obtain $\Acp$ and $\Afb$. Fitting the $\Acp$ values in bins of $\cosths$ to a constant, we obtain $\Acp = [-2.51\,\pm 1.44\,({\rm stat})\,^{+0.35}_{-0.52}\,({\rm syst})]\%$, where the first uncertainty is statistical and second is systematic \cite{Belle:2022gjv}.
\section{Measurement of $\acp$}
To measure $\acp$, we divide the data into four subsamples:
$D^0$ decays with $C^{}_T>0$ (${\rm yield}\!=\!N^{}_1$) and
$C^{}_T<0$ (${\rm yield}\!=\!N^{}_2$); and 
$\dbar$ decays with $-\overline{C}^{}_T>0$ ($N^{}_3$) and
$-\overline{C}^{}_T<0$ ($N^{}_4$).
Thus, $A^{}_T = (N^{}_1-N^{}_2)/(N^{}_1+N^{}_2)$,
$\bar{A}^{}_T = (N^{}_3-N^{}_4)/(N^{}_3+N^{}_4)$, and 
$\acp = (A^{}_T - \bar{A}^{}_T)/2$.
We fit the four subsamples simultaneously and take the 
fitted parameters to be $N^{}_1$, $N^{}_3$, $A^{}_T$, and~$\acp$. 
The fit gives  $\acp= [-1.95 \pm 1.42\,({\rm stat})\,^{+0.14}_{-0.12}\,({\rm syst})]\%$, where the first uncertainty is statistical and second is systematic \cite{Belle:2022gjv}.
\vspace{1cm}
\section{Conclusion:}
In summary, using 922~fb$^{-1}$ of Belle data we report the world's most precise branching fraction measurement for $\Dkskspp$ decays. The branching fraction, measured relative to that for $\Dkspp$, is:
$ \mathcal{B}(\Dkskspp)$/$\mathcal{B}(\Dkspp)  =  
[1.72\pm 0.03\,({\rm stat})\pm \textcolor{Black}{0.04}\,({\rm syst})\,]\times 10^{-2} $. Inserting the world average value $\mathcal{B}(\Dkspp) = (2.80\pm 0.18)\%$~\cite{Zyla:2020zbs} gives $\mathcal{B}(\Dkskspp) = 
 [4.82\pm 0.08\,({\rm stat})\pm ^{+0.10}_{-0.11}\,({\rm syst})\pm 0.31\,({\rm norm})]\times 10^{-4}  $ where the last uncertainty is due to $\mathcal{B}(\Dkspp)$. \\ 
 \indent We report the first $\cp$ violation search for $\Dkskspp$ using $\Acp$ and $\acp$. The time-integrated \cp\ asymmetry is measured to be $ \Acp(\Dkskspp) = [-2.51\pm 1.44\,({\rm stat})\, ^{+0.35}_{-0.52}\,({\rm syst})]$ \\ 
\indent The $\cp$-violating asymmetry $\acp$ is measured to be
$\acp(\Dkskspp) =  [-1.95\,\pm 1.42\,({\rm stat})\,^{+0.14}_{-0.12}\,({\rm syst})]$. 
Both $\Acp$ and $\acp$ measurements are consistent with zero $\cp$ violation.

\bigskip 
\bibliography{references}

\begin{thebibliography}{17}
\expandafter\ifx\csname natexlab\endcsname\relax\def\natexlab#1{#1}\fi
\expandafter\ifx\csname bibnamefont\endcsname\relax
  \def\bibnamefont#1{#1}\fi
\expandafter\ifx\csname bibfnamefont\endcsname\relax
  \def\bibfnamefont#1{#1}\fi
\expandafter\ifx\csname citenamefont\endcsname\relax
  \def\citenamefont#1{#1}\fi
\expandafter\ifx\csname url\endcsname\relax
  \def\url#1{\texttt{#1}}\fi
\expandafter\ifx\csname urlprefix\endcsname\relax\def\urlprefix{URL }\fi
\providecommand{\bibinfo}[2]{#2}
\providecommand{\eprint}[2][]{\url{#2}}

\bibitem[{\citenamefont{Allahverdi et~al.}(2021)}]{Allahverdi_2021}
\bibinfo{author}{\bibfnamefont{R.}~\bibnamefont{Allahverdi}}
  \bibnamefont{et~al.}, \bibinfo{journal}{Open Jour. Astrophys.}
  \textbf{\bibinfo{volume}{4}} (\bibinfo{year}{2021}).

\bibitem[{\citenamefont{Canetti et~al.}(2012)\citenamefont{Canetti, Drewes, and
  Shaposhnikov}}]{Canetti:2012zc}
\bibinfo{author}{\bibfnamefont{L.}~\bibnamefont{Canetti}},
  \bibinfo{author}{\bibfnamefont{M.}~\bibnamefont{Drewes}}, \bibnamefont{and}
  \bibinfo{author}{\bibfnamefont{M.}~\bibnamefont{Shaposhnikov}},
  \bibinfo{journal}{New J. Phys.} \textbf{\bibinfo{volume}{14}},
  \bibinfo{pages}{095012} (\bibinfo{year}{2012}).

\bibitem[{\citenamefont{Farrar and Shaposhnikov}(1994)}]{Farrar:1993hn}
\bibinfo{author}{\bibfnamefont{G.~R.} \bibnamefont{Farrar}} \bibnamefont{and}
  \bibinfo{author}{\bibfnamefont{M.~E.} \bibnamefont{Shaposhnikov}},
  \bibinfo{journal}{Phys. Rev. D} \textbf{\bibinfo{volume}{50}},
  \bibinfo{pages}{774} (\bibinfo{year}{1994}).

\bibitem[{\citenamefont{Sakharov}(1967)}]{Sakharov:1967dj}
\bibinfo{author}{\bibfnamefont{A.~D.} \bibnamefont{Sakharov}},
  \bibinfo{journal}{Pisma Zh. Eksp. Teor. Fiz.} \textbf{\bibinfo{volume}{5}},
  \bibinfo{pages}{32} (\bibinfo{year}{1967}).

\bibitem[{\citenamefont{Huet and Sather}(1995)}]{Huet:1994jb}
\bibinfo{author}{\bibfnamefont{P.}~\bibnamefont{Huet}} \bibnamefont{and}
  \bibinfo{author}{\bibfnamefont{E.}~\bibnamefont{Sather}},
  \bibinfo{journal}{Phys. Rev. D} \textbf{\bibinfo{volume}{51}},
  \bibinfo{pages}{379} (\bibinfo{year}{1995}).

\bibitem[{cha()}]{charge-conjugates}
\bibinfo{note}{Charge-conjugate modes are implicitly included unless noted
  otherwise.}

\bibitem[{\citenamefont{Collaboration}(2022)}]{Belle:2022gjv}
\bibinfo{author}{\bibfnamefont{B.}~\bibnamefont{Collaboration}}
  (\bibinfo{collaboration}{Belle Collaboration}) (\bibinfo{year}{2022}),
  \eprint{arXiv:hep-ex/2207.07555}.

\bibitem[{\citenamefont{Zyla et~al.}(2020)}]{Zyla:2020zbs}
\bibinfo{author}{\bibfnamefont{P.}~\bibnamefont{Zyla}} \bibnamefont{et~al.}
  (\bibinfo{collaboration}{Particle Data Group}), \bibinfo{journal}{PTEP}
  \textbf{\bibinfo{volume}{2020}}, \bibinfo{pages}{083C01}
  (\bibinfo{year}{2020}).

\bibitem[{\citenamefont{Grossman et~al.}(2007)\citenamefont{Grossman, Kagan,
  and Nir}}]{Grossman:2006jg}
\bibinfo{author}{\bibfnamefont{Y.}~\bibnamefont{Grossman}},
  \bibinfo{author}{\bibfnamefont{A.~L.} \bibnamefont{Kagan}}, \bibnamefont{and}
  \bibinfo{author}{\bibfnamefont{Y.}~\bibnamefont{Nir}},
  \bibinfo{journal}{Phys. Rev. D} \textbf{\bibinfo{volume}{75}},
  \bibinfo{pages}{036008} (\bibinfo{year}{2007}).

\bibitem[{\citenamefont{Aaij et~al.}(2019)}]{LHCb:2019hro}
\bibinfo{author}{\bibfnamefont{R.}~\bibnamefont{Aaij}} \bibnamefont{et~al.}
  (\bibinfo{collaboration}{LHCb Collaboration}), \bibinfo{journal}{Phys. Rev.
  Lett.} \textbf{\bibinfo{volume}{122}}, \bibinfo{pages}{211803}
  (\bibinfo{year}{2019}).

\bibitem[{\citenamefont{Durieux and Grossman}(2015)}]{Durieux:2015zwa}
\bibinfo{author}{\bibfnamefont{G.}~\bibnamefont{Durieux}} \bibnamefont{and}
  \bibinfo{author}{\bibfnamefont{Y.}~\bibnamefont{Grossman}},
  \bibinfo{journal}{Phys. Rev. D} \textbf{\bibinfo{volume}{92}},
  \bibinfo{pages}{076013} (\bibinfo{year}{2015}).

\bibitem[{\citenamefont{Valencia}(1989)}]{PhysRevD.39.3339}
\bibinfo{author}{\bibfnamefont{G.}~\bibnamefont{Valencia}},
  \bibinfo{journal}{Phys. Rev. D} \textbf{\bibinfo{volume}{39}},
  \bibinfo{pages}{3339} (\bibinfo{year}{1989}).

\bibitem[{\citenamefont{Bensalem and London}(2001)}]{Bensalem:2000hq}
\bibinfo{author}{\bibfnamefont{W.}~\bibnamefont{Bensalem}} \bibnamefont{and}
  \bibinfo{author}{\bibfnamefont{D.}~\bibnamefont{London}},
  \bibinfo{journal}{Phys. Rev. D} \textbf{\bibinfo{volume}{64}},
  \bibinfo{pages}{116003} (\bibinfo{year}{2001}).

\bibitem[{\citenamefont{Abashian et~al.}(2002)}]{Abashian:2000cg}
\bibinfo{author}{\bibfnamefont{A.}~\bibnamefont{Abashian}} \bibnamefont{et~al.}
  (\bibinfo{collaboration}{Belle Collaboration}), \bibinfo{journal}{Nucl.
  Instrum. Meth. A} \textbf{\bibinfo{volume}{479}}, \bibinfo{pages}{117}
  (\bibinfo{year}{2002}), \bibinfo{note}{also see Section 2 in J.~Brodzicka
  {\it et al.}, Prog. Theor. Exp. Phys. {\bf 2012}, 04D001 (2012).}

\bibitem[{\citenamefont{Kurokawa and Kikutani}(2003)}]{kekb}
\bibinfo{author}{\bibfnamefont{S.}~\bibnamefont{Kurokawa}} \bibnamefont{and}
  \bibinfo{author}{\bibfnamefont{E.}~\bibnamefont{Kikutani}},
  \bibinfo{journal}{Nucl. Instrum. Meth. A} \textbf{\bibinfo{volume}{499}},
  \bibinfo{pages}{1} (\bibinfo{year}{2003}), \bibinfo{note}{and other papers in
  this volume. T.~Abe {\it et al.}, Prog. Theor. Exp. Phys. {\bf 2013}, 03A001
  (2013) and references therein.}

\bibitem[{com()}]{comment_range}
\bibinfo{note}{\textcolor{Black}{ The fitted ranges are larger for the signal
  mode in order to more accurately model the background level.}}

\bibitem[{\citenamefont{Berends et~al.}(1973)\citenamefont{Berends, Gaemers,
  and Gastmans}}]{Afb}
\bibinfo{author}{\bibfnamefont{F.}~\bibnamefont{Berends}},
  \bibinfo{author}{\bibfnamefont{K.}~\bibnamefont{Gaemers}}, \bibnamefont{and}
  \bibinfo{author}{\bibfnamefont{R.}~\bibnamefont{Gastmans}},
  \bibinfo{journal}{Nucl. Phys.} \textbf{\bibinfo{volume}{B63}},
  \bibinfo{pages}{381} (\bibinfo{year}{1973}).

\end{thebibliography}
\end{document}